\documentclass[10pt]{article}
\usepackage[cp1251]{inputenc}
\usepackage[dvips]{graphicx}
\usepackage[dvips]{epsfig}

\def\lsim{\  \lower-1.2pt\vbox{\hbox{\rlap{$<$}\lower5pt\vbox{\hbox{$\sim$}}}}\ }
\def\gsim{\  \lower-1.2pt\vbox{\hbox{\rlap{$>$}\lower5pt\vbox{\hbox{$\sim$}}}}\ }

\begin{document}
\title{Bose-Einstein condensation in a one-dimensional system \\ of interacting bosons}
\author{ {\small Maksim Tomchenko}
\bigskip \\ {\small Bogolyubov Institute for Theoretical Physics, } \\
 {\small 14b,  Metrolohichna Str., 03680 Kyiv, Ukraine} \\
 {\small E-mail:mtomchenko@bitp.kiev.ua}}
 \date{\empty}
 \maketitle
 \large
 \sloppy

\begin{abstract}
 Using the Vakarchuk formulae
for the density matrix, we calculate the number $N_{k}$ of atoms with momentum $\hbar k$
for the ground state of a uniform one-dimensional periodic
system of interacting bosons. We obtain  for impenetrable point
bosons $ N_{0} \approx 2\sqrt{N}$ and $N_{k=2\pi j/L} \simeq 0.31N_{0}/\sqrt{|j|}$.
That is, there is no condensate or quasicondensate on low levels at large $N$.
For almost point bosons with weak coupling
($\beta=\frac{\nu_{0}m}{\pi^{2}\hbar^{2}n} \ll 1$), we obtain
$\frac{N_{0}}{N}  \approx   \left (\frac{2}{N\sqrt{\beta}}\right )^{\sqrt{\beta}/2} $ and $
N_{k=2\pi j/L} \approx  \frac{N_{0}\sqrt{\beta}}{4|j|^{1-\sqrt{\beta}/2}}$.
In this case, the quasicondensate exists on the level with $k=0$ and on low levels with $k\neq 0$,
if $N $ is large and  $ \beta$ is small
(e.g., for $N \sim 10^{10} $, $ \beta \sim 0.01$).
A method of measurement of such fragmented quasicondensate is proposed.
\end{abstract}

 \section{Introduction}
In the present work, we will study the Bose--Einstein condensation \cite{einstein1925}
for the ground state of a uniform one-dimensional (1D) periodic system of particles
with repulsive interaction. In some works, it is asserted that the condensate does not exist in the one-dimensional case.
This assertion is true only for infinite systems.
But all  systems in the Nature are finite. For the finite systems,
the macroscopic occupation of the one-particle state is possible, and it corresponds to a condensate \cite{einstein1925}.
The Bose--Einstein condensation in the momentum space depends on the behavior of the one-particle
density matrix $F_{1}(R=|\textbf{r}-\textbf{r}^{\prime}|)$
\cite{penronz}, which is the one-particle correlation function. If
the function $F_{1}(R)$ approaches a nonzero constant for large
$R,$ then the occupation number $N_{0}$ of the lowest one-particle
level is of the order of magnitude of the total number of atoms $N,$ and we
arrive at the condensate. If $F_{1}(R)$ slowly decreases (by a
power law or logarithmically), then the macroscopic occupation of the
one-particle state is possible. To distinguish this case from the
first one, it is accepted to talk about a quasicondensate
\cite{petrov2004,pethick2008}. For the fast (e.g., exponential)
decrease of $F_{1}(R),$ the macroscopic occupation of the
one-particle state is impossible; therefore, there is no
condensate or quasicondensate. In the 3D case, the states with
condensate and without condensate are possible. In 1D and 2D
cases, the quasicondensate is possible additionally; and, as usual, namely
the quasicondensate is realized instead of a ``true'' condensate.
We will consider impenetrable point bosons and almost point bosons with weak coupling.
In these extreme cases of the strong and weak interactions, the
wave functions of the ground state have the same structure.

It was shown in a series of works \cite{mw1966,hohenberg1967,kk1967,rc1967,pethick2008}
that, at a nonzero temperature
and $N, L \rightarrow \infty,$ the condensate on the level
with $k=0$ is forbidden for the 1D systems.
We will consider the case $T=0$, for which the behavior of the one-particle density matrix $F_{1}(R)$
was determined and it was shown that, in the limit $N, L \rightarrow \infty,$ the condensate is absent
\cite{lenard1964,popov1980,schwartz1977,vaidya1979,haldane1981,berkovich1989,petrov2003,forrester2003,mora2003,petrov2004,pethick2008}.
We will carry out the analysis on the basis of the Vakarchuk formulae for the density matrix \cite{vak1989,vak1990}.
In a similar approach, the analysis was executed in work \cite{schwartz1977},
but we will use a more accurate formula for the density matrix.  We will obtain known results and several new ones.

\section{Regime of infinitely strong coupling}
Consider the system of $N$ impenetrable point bosons
located on the periodic interval $[0,L]$.
The wave function of the ground
state of such system reads \cite{girardeau1960}
\begin{equation}
 \Psi_{0}=C \exp{\left (\frac{1}{2}\sum\limits_{j,l=1}^{N\prime} \ln{|\sin{[\pi(x_{j}-x_{l})/L]}|}\right )},
      \label{2} \end{equation}
where  $C=const$, and the prime above the sum means $j\neq l$.
Using the collective variables $\rho_{k} = \frac{1}{\sqrt{N}}\sum\limits_{j=1}^{N}e^{-ikx_j}$
and the expansion in the Fourier series
\begin{equation}
\frac{1}{2} \ln{|\sin{[\pi(x_{p}-x_{l})/L]}|} = \frac{1}{L}
 \sum\limits_{k_{j}}^{(2\pi)}\bar{\lambda}_{j}e^{ik_{j}(x_{p}-x_{l})},
     \label{6} \end{equation}
\begin{eqnarray}
&&\bar{\lambda}_{j} = \frac{1}{2}\int\limits_{0}^{L}dx
\ln{[\sin{(\pi x/L)}]}e^{-ik_{j}x},
 \label{7} \end{eqnarray}
we can write the function $\Psi_{0}$ (\ref{2}) in the form \cite{girardeau1960}
\begin{equation}
 \Psi_{0}=C^{\prime}e^{\frac{1}{2}\sum\limits_{k\neq 0}^{(2\pi)}a_{2}(k)\rho_{k}\rho_{-k}},
      \label{9} \end{equation}
 where
\begin{equation}
 a_{2}(k_{j})=2N\lambda_{j}, \quad C^{\prime}=C e^{N^{2}\lambda_{0}-N\sum\limits_{j}\lambda_{j}},
      \label{10} \end{equation}
\begin{eqnarray}
&&\lambda_{j} =\bar{\lambda}_{j}/L = \frac{1}{2}\int\limits_{0}^{1}dt
\ln{[\sin{(\pi t)}]}\cos{(2\pi jt)}.
 \label{8} \end{eqnarray}
Here and below, the symbol $(l\pi)$ above the sum means that $k_{j}$ runs the values
$k_{j}=l\pi j/L$, $j=0, \pm 1, \pm 2, \ldots$
Since $\int\limits_{0}^{L}dx \ln{[\sin{(\pi x/L)}]}\sin{(k_{j}x)}=0$, we
write in (\ref{8}) $\cos{(2\pi jt)}$ instead of $e^{-i2\pi jt}$.
It was found  \cite{girardeau1960} that
\begin{eqnarray}
\lambda_{0}=-\frac{\ln{2}}{2}, \quad  \lambda_{j\neq 0} =-\frac{1}{4|j|}.
 \label{lam} \end{eqnarray}
It can be proved by the direct numerical calculation that formulae (\ref{lam}) are proper, and series (\ref{6}), (\ref{7})
restores the function $(1/2) \ln{|\sin{[\pi(x_{p}-x_{l})/L]}|}$ exactly.

  I. Vakarchuk \cite{vak1989} developed a method of calculation of the $s$-particle density matrix,
which for the ground state reads
\begin{eqnarray}
 &&F_{s}(\textbf{r}_{1},\ldots,\textbf{r}_{s}|\textbf{r}_{1}^{\prime},\ldots,\textbf{r}_{s}^{\prime}) =
V^{s} \int d\textbf{r}_{s+1}\ldots d\textbf{r}_{N} \Psi^{*}_{0}(\textbf{r}_{1}^{\prime},\ldots,\textbf{r}_{s}^{\prime},\textbf{r}_{s+1},\ldots , \textbf{r}_{N})
 \times \nonumber \\ &\times & \Psi_{0}(\textbf{r}_{1},\ldots,\textbf{r}_{s},\textbf{r}_{s+1},\ldots ,\textbf{r}_{N}).
      \label{11} \end{eqnarray}
For $\Psi_{0}$ of the form (\ref{9}), the formulae from \cite{vak1989} yield the following series
for the logarithm of the one-particle density matrix \cite{vak1990}:
\begin{equation} \ln{F_{1}(x,x^{\prime})} = u_{1}(R) + u_{2}(R) + \ldots, \quad R=x-x^{\prime},
      \label{13} \end{equation}
\begin{equation}
u_{1}(R) =\frac{1}{N}\sum\limits_{k\neq 0}^{(2\pi)}\frac{a_{2}^{2}(k)}{1-2a_{2}(k)}
\left (e^{ikR}-1 \right ),
      \label{14} \end{equation}
\begin{equation}
u_{2}(R) =\frac{1}{N^{2}}\sum\limits_{k_{1},k_{2}}^{(2\pi)}
\frac{a_{2}(k_{1})a_{2}(k_{2})a_{2}(-k_{1}-k_{2})}{(1-2a_{2}(k_{1}))^{2}(1-2a_{2}(k_{2}))(1-2a_{2}(-k_{1}-k_{2}))}
\left (e^{ik_{1}R}-1 \right ).
      \label{15} \end{equation}
In sum (\ref{15}), $k_{1},k_{2},k_{1}+k_{2}\neq 0$.
Two last formulae are true for large $N, L$.

 The analysis was performed on the basis of the density matrix also in work \cite{schwartz1977}.
In the approximation of small fluctuations of the density and the current,
the following formulae \cite{schwartz1974} were obtained:
\begin{equation}
F_{1}(R)|_{T=0} =e^{\tilde{u}(R)}\left [1-\frac{1}{2N}\sum\limits_{p\neq 0}^{(2\pi)}(S_{p}-1)(1-e^{ipR})\right ],
      \label{14s} \end{equation}
\begin{equation}
\tilde{u}(R) =\frac{1}{N}\sum\limits_{k\neq 0}^{(2\pi)}\frac{S_{k}^{2}-1}{4S_{k}}(1-\cos{kR}).
      \label{15s} \end{equation}
In order to compare formulae (\ref{14s}) and (\ref{15s}) with (\ref{13})--(\ref{15}), we note the following.
For a system of interacting bosons with any \textit{finite} coupling constant (penetrable particles),
$ \Psi_{0}$ takes the form \cite{yuv1980}
\begin{equation}
 \Psi_{0}=C \exp{\left(\frac{1}{2!}\sum\limits_{k\neq 0}^{(2\pi)}a_{2}(k)\rho_{k}\rho_{-k}+
 \frac{1}{3!}\sum\limits_{k_{1},k_{2}\neq 0}^{(2\pi)\prime}a_{3}(k_{1},k_{2})\rho_{k_{1}}\rho_{k_{2}}\rho_{-k_{1}-k_{2}}+\ldots \right ) },
      \label{31} \end{equation}
where the prime above the sum means $k_{1}+k_{2}\neq 0$. The analysis \cite{yuv1980} is valid,
generally speaking, for nonpoint particles. For point penetrable bosons,
we have the Lieb--Liniger solution for $\Psi_{0}$ \cite{ll1963}. Apparently, the analysis
in \cite{yuv1980} is also proper for the point particles,
so that $\Psi_{0}$ given by the Lieb--Liniger solution can be written in the form
(\ref{31}). But this question was not considered in the literature, to our knowledge (see also \cite{pointbosonsme2015}).
For the point bosons with infinite positive coupling constant (impenetrable bosons), the solution has the
form (\ref{9}), which follows from (\ref{31}) provided  $a_{j\geq 3}= 0$.
For the penetrable nonpoint bosons, $a_{j\geq 3}\neq 0$.
Thus, the penetrable nonpoint bosons and the impenetrable point ones can be described in a unified way,
by starting from $\Psi_{0}$ (\ref{31}).

For nonpoint bosons in the regime of weak coupling, the relation
$2a_{2}(k)\approx 1-1/S_{k}$ holds \cite{yuv1980}, and, for not too small $k,$ the quantity $a_{2}(k)$ is small.
We can verify that, in this case, the sum on the right-hand side of (\ref{14s})
is small, and it can be raised in the exponent.
Then (\ref{14s}) and (\ref{15s})
are reduced to $F_{1}(R)=e^{u_{1}(R)}$ with $u_{1}$ (\ref{14}).
That is, the density matrix from \cite{schwartz1977} at a weak coupling
coincides with the first approximation for the density matrix (\ref{13})--(\ref{15}) \cite{vak1989}.
It is possible to restrict oneself  in Eqs. (\ref{13})--(\ref{15}) to the first approximation
($u_{1}\neq 0, \ u_{2}=0$), if the coupling is weak
(see the following section). It follows that the approximation of small fluctuations \cite{schwartz1977,schwartz1974}
is equivalent to the approximation of weak coupling. Moreover,
the Feynman formula $S_{k}=\hbar^{2}k^{2}/2mE(k)$ was used  in \cite{schwartz1977}.
For a weak coupling, this formula is close to the exact one for all $k$;
for a strong coupling, it is valid only for small $k$.
The impenetrable point bosons correspond to the infinitely strong coupling.
In this case, $a_{2}(k)$ in Eq. (\ref{31}) is set by formulae  (\ref{10})
and (\ref{8}), and $a_{j\geq 3}= 0$.   In this case, the density matrices
(\ref{13})--(\ref{15}) and (\ref{14s}), (\ref{15s}) do not coincide with one another.
In work \cite{vak1989}, the perturbation theory is constructed for the logarithm
of the density matrix, and it is valid for any coupling
(the results given below indicate that, even for a strong coupling, series
(\ref{13}) is apparently rapidly convergent).
Thus, formulae (\ref{13})--(\ref{15}) \cite{vak1989,vak1990} are more accurate than formulae
(\ref{14s}) and (\ref{15s})  \cite{schwartz1977,schwartz1974}, because  the former involve the
following correction.
In addition, some constants were not determined in \cite{schwartz1977}. We will find all the constants.
In these respects, our analysis is better than the analysis  \cite{schwartz1977}.

With regard for (\ref{10}) and the equalities
$\lambda_{j}=\lambda_{-j}$, $k_{j}=2\pi j/L,$ formulae (\ref{14}) and (\ref{15}) can be written in the form
\begin{equation}
u_{1}(R) =\sum\limits_{j=\pm 1, \pm 2, \ldots}\frac{4N\lambda_{j}^{2}}{1-4N\lambda_{j}}
\left (e^{i2\pi jR/L}-1 \right ),
      \label{16} \end{equation}
\begin{equation}
u_{2}(R) =\sum\limits_{j_{1},j_{2}=\pm 1, \pm 2, \ldots}^{j_{1}+j_{2}\neq 0}
\frac{8N\lambda_{j_{1}}\lambda_{j_{2}}\lambda_{j_{1}+j_{2}}}{(1-4N\lambda_{j_{1}})^{2}(1-4N\lambda_{j_{2}})(1-4N\lambda_{j_{1}+j_{2}})}
\left (e^{i2\pi j_{1}R/L}-1 \right ).
      \label{17} \end{equation}
The average number of particles with momentum
$\hbar k$ in the 1D case is determined  by the well-known formula
\begin{equation}
N_{k} =\frac{N}{L}\int\limits_{0}^{L}F_{1}(R)e^{-ikR} dR.
      \label{18} \end{equation}
Such approach allows one to obtain the reasonable estimates for the condensate in He II \cite{vak1990,me2006}.

In the literature, the condensate is frequently defined by the formula \cite{penronz}
\begin{equation}
N_{0} =NF_{1}(R\rightarrow \infty),
      \label{18b2} \end{equation}
which is true in the thermodynamical limit ($N, V \rightarrow \infty, N/V=const)$.
We now consider 1D periodic systems of \textit{finite} size $L$.
The periodicity yields $F_{1}(0)=F_{1}(L)$. The analysis below indicates
that the density matrix $F_{1}(R)$ takes the maximum value ($F_{1}=1$) at the ends of the interval ($R=0, L$)
and decreases, while approaching the middle of the interval.  The quantity $F_{1}(R)$ is minimal for $R= L/2$.
Therefore, for the finite periodic 1D systems, formula (\ref{18b2}) should be replaced by
\begin{equation}
N_{0} \approx NF_{1}(R\rightarrow L/2).
      \label{18c2} \end{equation}
 Formula (\ref{18c2}) underestimates $N_{0}$
as compared with the exact value (\ref{18}), which is evident (i) for a strong coupling or (ii) for
small $N$ in the case of weak coupling.

 Using formulae
(\ref{18}), (\ref{13}), (\ref{16}), and (\ref{17}), we find now the values of $N_{k}$ for the
ground state of $N$
impenetrable point bosons in a cyclic vessel by means of a direct numerical summation.
Since sums (\ref{16}), (\ref{17})  are present in (\ref{18}) in the exponent, we need to take rather many terms
($\gsim 10^{6}$ for the summation over each $j$) in order to attain a good accuracy in sums (\ref{16}) and (\ref{17}).
The results for $N=10^2$--$10^4$ are as follows:
\begin{equation}
N_{0}= C_{1}\sqrt{N},
      \label{19} \end{equation}
\begin{equation}
N_{k=2\pi l/L} = C_{2}N_{0}/\sqrt{|l|}, \quad 1\leq |l|\ll N,
      \label{19b} \end{equation}
where $C_{1}=0.87\pm 0.01, C_{2}=0.33\pm 0.005 $ in the first approximation, and
$C_{1}=1.99\pm 0.05, C_{2}=0.31\pm 0.03$ in the second one
(we take only $u_{1}$ into account in (\ref{13}) in the
first approximation and $u_{1}$, $u_{2}$ in the second one).

Estimates (\ref{19}) and (\ref{19b})  can be obtained analytically for the first approximation.
Using formulae (\ref{16}) and (\ref{lam}), we write the function $u_{1}(R)$ in the form
\begin{equation}
u_{1}(R) =\sum\limits_{j= 1, 2, \ldots} \alpha_{j}(\cos{(2\pi jR/L)}-1), \quad \alpha_{j}=\frac{N}{2j(N+j)}.
      \label{21} \end{equation}
In formula (\ref{18}) for $N_{k},$ the function
$F_{1}(R)=e^{u_{1}(R)}$ stands under the sign of integral. It
follows from formula (\ref{50})  below that the value of
$|u_{1}(R)|$ is usually large: for example, for $N=3\cdot 10^6,$ we
have $|u_{1}(R)|\lsim 8$. Therefore, it is not expedient to expand
$e^{u_{1}(R)}$ in a series, since too many terms should be taken
into account in order to obtain a proper result. It is better to
determine the exponent in $F_{1}(R)=e^{u_{1}(R)}.$ Relation
(\ref{21}) and the Euler--Maclaurin formula
\begin{equation}
\sum\limits_{j=1, 2, \ldots}f(j)\approx \int\limits_{1}^{\infty }f(x)dx+B_{1}(f(\infty)-f(1))+
\frac{B_{2}}{2}(\acute{f}(\infty)-\acute{f}(1))
     \label{23} \end{equation}
with the Bernoulli numbers $B_{1}=-1/2$ and $B_{2}=1/6$ yield for $1/N\lsim R/L \leq 1/2$:
   \begin{equation}
F_{1}(R) \approx e^{u_{1}(R)}\approx \left (\frac{f(R)}{nR}\right )^{1/2}, \quad
  f(R)\approx 0.089+0.2(R/L)^{2},
      \label{50} \end{equation}
where $n=N/L.$ The fitting function $f(R)$ was determined by means of the comparison of $u_{1}(R)$ with
the results of a numerical summation of (\ref{21}). This function allows one to get the numerical values of
$u_{1}(R)$ for $R=0.001L$--$0.5L$ and $N=10^3$--$10^6$ with a small error of $\lsim 0.2\% $.
For $R> L/2,$ it is necessary to change $R\rightarrow  L - R$
on the right-hand side of (\ref{50}). Then relation (\ref{50}) yields
\begin{equation}
N_{k=2\pi l/L} =\frac{N}{L}\int\limits_{0}^{L}F_{1}(R)e^{-ikR} dR \approx
2\sqrt{N}\int\limits_{0}^{1/2}dt\cos{(2\pi l t)}\sqrt{\frac{0.089}{t}+0.2t}.
      \label{18b} \end{equation}
This gives formula (\ref{19}) for $N_{0}$ with constant $C_{1}\approx 0.89$, which is close to the above-given
value $C_{1}\approx 0.87$. Integral (\ref{18b})
can be easily found numerically, and, for any $l\neq 0,$ the answer is as follows:
\begin{equation}
N_{k=2\pi l/L } = (0.295\pm 0.003)\sqrt{N/|l|}\approx 0.331N_{0}/\sqrt{|l|}.
      \label{18c} \end{equation}
If we eliminate the term $0.2t$ from (\ref{18b}), then the law
  $N_{k=2\pi l/L }\sim 1/\sqrt{|l|}$ is satisfied for small $|l|$ with less accuracy.
For $N=400,$ formula (\ref{18c}) gives the value, which is overestimated by $ 10\% $ relative to
the result of a direct numerical  summation in (\ref{18}), (\ref{21}).
But, as $N$ increases, this difference decreases to $5\% $ for $N=2000$ and to $1 \% $
for $N=10^{4}.$

The following results were obtained previously. For $L=N,$ it was shown \cite{lenard1964} that
\begin{equation}
N_{0} <  2\sqrt{e N}, \quad F_{1}(|R|\rightarrow \infty)|_{N\rightarrow \infty} \leq  (e/\pi |R|)^{1/2},
      \label{501} \end{equation}
which agrees with (\ref{19}) and (\ref{50}).
The dependence $F_{1}(|R|\gg n^{-1}) \sim |R|^{-1/2}$ was found in \cite{schwartz1977}.
Formula (\ref{50}) with $f(R)= 1$ was deduced in \cite{petrov2004}.
The formulae $F_{1}(|R|\gg n^{-1}) \sim |R|^{-1/2}, N_{k\neq 0}\sim |k|^{-1/2}$ were gotten in \cite{vaidya1979}.
The  exact calculation \cite{forrester2003} gives
  \begin{equation}
F_{1}(R) = \frac{0.924}{\sqrt{N\sin{(\pi R/L)}}}\approx \left (\frac{0.27+0.45(R/L)^2}{nR}\right )^{1/2},
      \label{for1} \end{equation}
\begin{equation}
N_{0}\approx 1.543\sqrt{N}, \quad  N_{k=2\pi l/L } \approx \frac{0.338N_{0}}{\sqrt{|l|}} \quad (l\neq 0).
      \label{for2} \end{equation}
In our approach, the direct numerical summation in (\ref{16})--(\ref{18}) in the second approximation
gives $N_{k}$ (\ref{19}), (\ref{19b}) and the density matrix
 \begin{equation}
F_{1}(R) = e^{u_{1}(R)+u_{2}(R)}\approx \left (\frac{e^{1.64}f(R)}{nR}\right )^{1/2}\approx
    \left (\frac{0.46+(R/L)^2}{nR}\right )^{1/2}.
      \label{502} \end{equation}
This is in approximate agreement with the results
\cite{schwartz1977,vaidya1979,petrov2004,forrester2003}.

It is seen from formulae (\ref{19}), (\ref{19b}),  (\ref{50}),  and (\ref{502})
that, in our approach, the results for $N_{0}$, $N_{k\neq 0 },$ and $F_{1}(R)$ in the second approximation
are approximately by a factor $2.3$ larger than in the first approximation.
Such significant difference is related to the absence of a small parameter in
expansion (\ref{13}) and to the fact that this is the expansion of the value in the exponent.
However, the results in the second approximation are in better agreement with the exact ones \cite{forrester2003}
(as compared with the results in the first approximation) and differ from the latter by $\leq 30\%$.
As for the ratio $N_{k}/N_{0},$ the method gives the result (\ref{19b}), which differs
from the exact one (\ref{for2}) only by $10\%$.
That is, the method allows one to get reasonable results, and we expect that, with the account for several following
$u_{j}$ in (\ref{13}), the results will be close to the exact ones.

It is of interest that, according to our analysis, $u_{1}$ depends
weakly on $R$, and $u_{2}(R)$ is a constant $0.82\pm 0.01$
everywhere except for narrow bands $|R|\lsim 1/N$ and $|L-R|\lsim
1/N$. In this case, relation $|u_{2}(R)|\simeq q|u_{1}(R)|$ holds,
where $q\approx 0.224\ln{(1925)}/\ln{(3.85N)}$ (e.g., $q\approx
0.22$ for $N=500$). If the same law of decrease of $|u_{j}(R)|$ with
increase in $j$ holds for the following  $j,$ then even the first
approximation (\ref{19b}) for $N_{k}/N_{0}$ should be close to the
exact value, and the second approximation (\ref{19}) for $N_{0}/N$
should differ from the exact value by at most several tens of
percents. The comparison of results (\ref{19}), (\ref{19b}) with the
exact solutions (\ref{for2}) confirms these properties. This allows
us to expect that, though our approach has no small parameter and
the correction $u_{2}(R)$ affects considerably the result, the
following corrections $u_{j\geq 3}$ will less affect  the results.

We note that the condition
\begin{equation}
\sum\limits_{k}N_{k}= N
     \label{29} \end{equation}
holds automatically. This is related to  that
$LN_{k}/N$ is the Fourier transform of the function $F_{1}(R)$, according to (\ref{18}). Therefore,
$F_{1}(0)=(1/L)\sum\limits_{k}LN_{k}/N$. In the first and second approximations, $F_{1}(0)=1$,
which yields (\ref{29}).
We note also that the function $F_{1}(R=x-x^{\prime})$  depends on two arguments
($x\in [0,L]$ and  $x^{\prime}\in [0,L]$) and is periodic in each argument with period $L$.
In this case, the equality $F_{1}(R)= F_{1}(|R|)$ holds.
Therefore, $F_{1}(|R|)$ can be expanded in a single Fourier series on the interval $|R|\in[0,L]$.
Formula (\ref{18}) sets the Fourier transform for such a series.
The same is true for expansion (\ref{6}), (\ref{7}), because $\ln{|\sin{(\alpha)}|}=\ln{|\sin{|\alpha}||}$.

\section{Regime of a weak coupling}
Consider an analogous problem for the ground state
of a 1D system with weak coupling (highly penetrable bosons).
To simplify the formulae, we consider the interatomic potential $U(x_{i}-x_{j})$
to be an extremely high narrow barrier
close to the $\delta$-function with the Fourier transform $\nu(k)=\nu_{0}=const$.
For a system of penetrable bosons, we have  $\Psi_{0}$ (\ref{31}).
Under a weak coupling (weak interaction $ \nu_{0}$
or a high concentration, $\beta \ll 1$ in (\ref{34})), the correction $a_{3}(k_{1},k_{2})$ in (\ref{31})
is small, and the sum with $a_{3}$ can be neglected \cite{yuv1980}. Therefore,
$\Psi_{0}$ takes the form (\ref{9}) with $a_{2}(k)$ to be \cite{bz1955}
\begin{equation}
2a_{2}(k)\approx 1-\sqrt{1+\frac{4n\nu_{0}m}{\hbar^{2}k^{2}}},
      \label{32} \end{equation}
and formulae (\ref{13})--(\ref{15}) remain valid. With regard for (\ref{32}), we get
\begin{equation}
u_{1}(R) =\sum\limits_{j= 1, 2, \ldots} \alpha^{p}_{j}(\cos{(2\pi jR/L)}-1),
      \label{33} \end{equation}
\begin{equation}
\alpha^{p}_{j}=\frac{1+\beta N^{2}/(2j^{2})-\sqrt{1+\beta N^{2}/j^{2}}}{N\sqrt{1+\beta N^{2}/j^{2}}},
\quad \beta=\frac{\nu_{0}m}{\pi^{2}\hbar^{2}n}.
      \label{34} \end{equation}
Here, $\beta $ is a dimensionless coupling constant.  $u_{2}(R)$ (\ref{15}) can be represented in the form
(\ref{33}) too. In this case, $\alpha^{p}_{j}$ is different and much less in modulus
(for $\beta \ll 1$). Therefore, the correction $u_{2}(R)$
can be neglected. For sufficiently small $\beta,$ $|u_{1}(R)|\ll 1$ is satisfied (see Eq. (\ref{40}) below).
Therefore, the exponential function $e^{u_{1}(R)}$ can be expanded in a series. Then we have
\begin{eqnarray}
&&\frac{N_{k=2\pi l/L}}{N} = L^{-1}\int\limits_{0}^{L} d R \cos{(2\pi l R/L)}\exp{[\sum\limits_{j= 1}^{\infty } \alpha^{p}_{j}(\cos{(2\pi jR/L)}-1)]}
\approx \label{27a} \\ &\approx &  e^{-\sum\limits_{j= 1}^{\infty } \alpha^{p}_{j}}
\int\limits_{0}^{1}dt \cos{(2\pi l t)}\left (1+ \sum\limits_{j= 1}^{\infty } \alpha^{p}_{j}\cos{(2\pi j t)}+\frac{1}{2!}
\sum\limits_{j_{1},j_{2}= 1}^{\infty } \alpha^{p}_{j_{1}}\alpha^{p}_{j_{2}}\cos{(2\pi j_{1}t)}\cos{(2\pi j_{2}t)}
\right ),
       \nonumber \end{eqnarray}
where $t=R/L$. This implies
\begin{eqnarray}
\frac{N_{0}}{N} \approx   e^{-\sum\limits_{j= 1}^{\infty } \alpha^{p}_{j}}
\left (1+ \frac{1}{4}\sum\limits_{j= 1}^{\infty } (\alpha^{p}_{j})^{2}\right )\approx   e^{-\sum\limits_{j= 1}^{\infty } \alpha^{p}_{j}},
      \label{22} \end{eqnarray}
\begin{eqnarray}
&&\frac{N_{k=2\pi l/L}}{N} \approx e^{-\sum\limits_{j= 1}^{\infty } \alpha^{p}_{j}}
\left (\frac{\alpha^{p}_{l}}{2}+\frac{1}{8}\sum\limits_{j= 1}^{l-1 } \alpha^{p}_{j}\alpha^{p}_{l-j}+
\frac{1}{4}\sum\limits_{j= 1}^{\infty } \alpha^{p}_{j}\alpha^{p}_{l+j}\right )\approx
\frac{\alpha^{p}_{l}}{2}e^{-\sum\limits_{j= 1}^{\infty } \alpha^{p}_{j}},
      \label{27} \end{eqnarray}
where $l=\pm 1, \pm 2, \ldots$. It can be verified that, for $|l|\ll N\sqrt{\beta},$ the modulus of each of the sums with $(\alpha^{p})^{2}$
in (\ref{22}), (\ref{27}) is less that the principal term ($1$ or $\alpha^{p}_{l}/2$) by $\sim \beta^{-1}$ or
$\sim \beta^{-1/2}$ times. Therefore, we can neglect these sums, if $\beta$ is small.
For $|l| \gsim 0.1N\sqrt{\beta},$ the rejected corrections ($\sim \alpha^{2}, \alpha^{3}, \ldots $) decrease
the value of $N_{k}$ significantly.
We obtain by formula  (\ref{23}) that, for $N\sqrt{\beta}\gg 1,$
\begin{eqnarray}
\sum\limits_{j= 1}^{\infty } \alpha^{p}_{j} \approx
\frac{\sqrt{\beta}}{2} \ln{(N\sqrt{\beta}/2)}.
      \label{37} \end{eqnarray}
Since the relation $\alpha^{p}_{l}\approx \frac{\sqrt{\beta}}{2|l|}$ holds for $|l|\ll N\sqrt{\beta},$ we finally have
\begin{eqnarray}
N_{0} \approx &  N\left (\frac{2}{N\sqrt{\beta}}\right )^{\sqrt{\beta}/2},
\quad N_{k=2\pi l/L} \approx  \frac{N_{0}\sqrt{\beta}}{4|l|},
      \label{38} \end{eqnarray}
where $1\leq |l|\ll N\sqrt{\beta}$.

The values of $N_{k}$ can be found in a different way. For
$(\sqrt{\beta}N)^{-1} \lsim R/L \leq 1/2$ and $\sqrt{\beta}N\gg 1$, we have
\begin{equation}
u_{1}(R) \approx \frac{\sqrt{\beta}}{2}\ln{\frac{f_{2}(R)}{\sqrt{\beta}nR}}, \quad f_{2}(R)\approx  \frac{0.98+2(R/L)^{2}}{3}.
      \label{39} \end{equation}
The main dependence in (\ref{39}) can be found by formula (\ref{23}), and the fitting function
 $f_{2}$ follows from the direct numerical summation of (\ref{33}).
From whence, we get the density matrix
   \begin{equation}
F_{1}(R) \approx e^{u_{1}(R)}\approx \left (\frac{f_{2}(R)}{\sqrt{\beta}nR}\right )^{\sqrt{\beta}/2}.
      \label{40} \end{equation}
For $R> L/2,$ we should replace $R\rightarrow  L - R$ on the
right-hand sides of (\ref{39}) and (\ref{40}).

In Fig. 1, we show the calculated density matrix. Two curves are
significantly different for $R/L\lsim 0.0001$ and $R/L\gsim 0.9999$.
For $0.0001 < R/L < 0.9999,$ the curves practically coincide: that
is, formula (\ref{40}) with $f_{2}(R)$ (\ref{39}) is very close to
the exact numerical solution for $F_{1}(R)$. The smallest value of
$F_{1}(R)$ is $F_{1}(R=L/2)\approx 0.384$.
\begin{figure}[ht]
\centerline{\includegraphics[width=85mm]{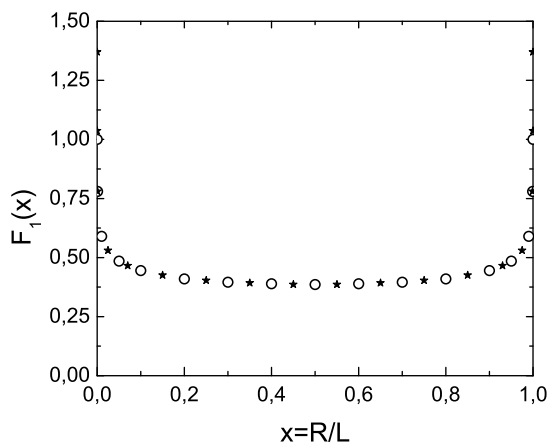} } \caption{
Density matrix $F_{1}(R)$ for a periodic 1D system of almost point
bosons with weak coupling ($N=10^{4}$, $\beta = 0.06$). The direct
numerical calculation of $F_{1}(R)$ on the basis of the exact
formulae (\ref{18}), (\ref{13}), (\ref{33}), (\ref{34}) with
$u_{j\geq 2}=0$ ($\circ\circ\circ$) and $F_{1}(R)$ (\ref{39}),
(\ref{40})  with the replacement $R\rightarrow  L - R$ for $R> L/2$
($\star\star\star$).
 \label{fig1}}
\end{figure}

Relations (\ref{18}) and (\ref{40}) yield
\begin{eqnarray}
N_{0} \approx &  NI_{0}\cdot\left (\frac{2}{N\sqrt{\beta}}\right
)^{\sqrt{\beta}/2}, \quad N_{k=2\pi l/L} \approx
N_{0}\frac{I_{l}}{I_{0}} \quad (l\neq 0),
      \label{41} \end{eqnarray}
\begin{equation}
I_{l} =I_{-l}= 2\int\limits_{0}^{1/2}dt\cos{(2\pi l t)}\left (\frac{0.49}{3t}+\frac{t}{3}\right )^{\sqrt{\beta}/2}.
      \label{42} \end{equation}
For $\beta \lsim 0.01$, we find numerically
\begin{equation}
I_{l=0}\equiv I_{0} \approx 1, \quad I_{l\neq 0} \approx \frac{\sqrt{\beta}}{4|l|^{1-\sqrt{\beta}/2}}.
      \label{43} \end{equation}
Formulae (\ref{41}) are close to (\ref{38}). For $\beta = 0.1,$
formulae (\ref{43}) underestimate the values of $I_{l\neq 0}$
(\ref{42}) by  $10 \% $ and $I_{0}$  (\ref{42}) by $2 \% $. Both
formulae (\ref{41}), (\ref{42}) and formulae (\ref{38})  agree with
the results of a direct numerical calculation of $N_{k}$ on the
basis of (\ref{18}), (\ref{13}), (\ref{33}), and
 (\ref{34}). But formulae  (\ref{41}) and (\ref{42})  are more accurate than (\ref{38}) (see below).

It follows from (\ref{39}) and (\ref{40}) that formula (\ref{38}) for $N_{0}$ can be written as
\begin{eqnarray}
N_{0} \approx   N\cdot 2^{\sqrt{\beta}/2}\cdot F_{1}(L/2).
      \label{38b} \end{eqnarray}

\begin{figure}[ht]
\centerline{\includegraphics[width=85mm]{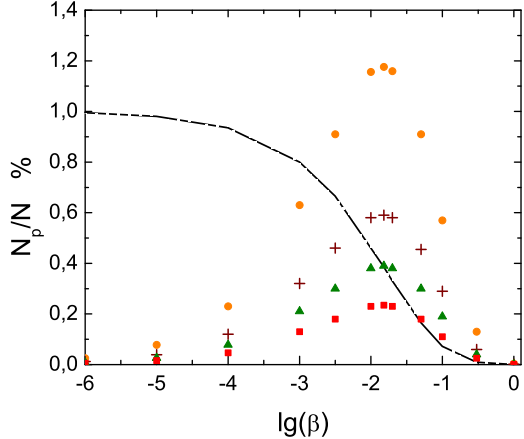} } \caption{
Fragmented quasicondensate. Values of $100N_{p}/N$ (Eqs. (\ref{38}))
for $N=10^8$ and $p=0$ (solid line), $p=1$ ($\circ\circ\circ$),
$p=2$ ($+++$), $p=3$ ($\bigtriangleup\bigtriangleup\bigtriangleup$),
$p=5$ (squares). $N_{0}$ is reduced by $100$ times.
 \label{fig2}}
\end{figure}

Formulae (\ref{38}), (\ref{41}) and (\ref{43}) indicate that, for
small $\beta$ and large $N,$ the quasicondensate is present not only
on the level with $k=0$, but also on low levels with $k\neq 0$ (see
Fig. 2; if $\beta$ is very small, then the quasicondensate occupies
only the level $k=0$). Such \textit{fragmented quasicondensate} is,
in some sense, a corroboration of M. Girardeau ideas
\cite{girardeau1960}, but for a weak coupling. It is of interest to
mention the work by E. Witkowska et al. \cite{witkowska2011}, where
a model of evaporative cooling of a one-dimensional gas in a trap
was constructed. It was found that several lower levels are
macroscopically filled in the initial nonequilibrium regime and only
the lowest level is macroscopically filled in the final equilibrium
regime (though, it was not explained in this work how the occupation
numbers are calculated; their determination is a complicated task:
it is necessary to find the density matrix and then to determine the
occupation numbers $\lambda_{j}$ from Eq. (\ref{46})). The results
\cite{witkowska2011} imply that the macroscopic occupation of
several lower levels is related namely to the absence of
equilibrium, i.e. to the disorder. This is not  quite clear
physically, since a disorder destroys the macroscopic occupation of
levels, as usual. Possibly, the effect is related to a comparatively
small $N$ ($N\leq 10^{4}$) and will disappear, as $N$ will increase
by at least two orders. The equilibrium state \cite{witkowska2011}
corresponds, probably, to small $\beta $: $\beta\leq 0.01$. In this
case, formula (\ref{38}) yields the macroscopic occupation for the
lowest level only. Above, we have found a fragmented quasicondensate
for a uniform equilibrium 1D system of interacting spinless bosons,
by exactly describing  the interaction. Apparently, such solution
was not obtained previously. Note that the regimes, in which a
generalized condensate appears in the ideal gas, were investigated
in \cite{mullin2012}. Several particular systems with possible
fragmented condensates were discussed in \cite{leggett2006}.

   According to (\ref{38}),
the occupation of low levels with $k\neq 0$ is maximal for $N=10^{10}$ and $\beta \approx 0.009$
($N_{0} \approx 0.388N$,  $N_{l} \approx 0.0092N/|l|$). For
$N=10^{4},$  the occupation is maximal for $\beta \approx 0.06$
($N_{0} \approx 0.419N$,  $N_{l} \approx 0.0256N/|l|$, see also Fig. 1).
Formulae (\ref{41}) and (\ref{43})  give practically the same  values. The numerical
calculation on the basis of the exact relations (\ref{18}), (\ref{13}), (\ref{33}), and
 (\ref{34}) for $N=10^{4}$, $\beta \approx 0.06$ gives $N_{0}$ to be by $1 \%$
 larger relative to (\ref{38}) and (\ref{41}), (\ref{43}) and $N_{k \neq 0}$ to be
by $10 \%$ larger relative to (\ref{38}) and  by $8 \%$ larger relative to (\ref{41}), (\ref{43}) (for $|l|\ll N\sqrt{\beta}$).
By comparing with the more nearly exact formulae (\ref{41}) and (\ref{42}), $N_{0}$ is only by $0.1 \%$ larger,
and $N_{k \neq 0}$ by $(1\div 2) \%$. As $\beta$ decreases, all these differences decrease as well.

Our results agree with those obtained earlier. From the study of the
density matrix \cite{schwartz1977} and
 from the study of fluctuations \cite{haldane1981,petrov2003,petrov2004,pethick2008}, the relation
 \begin{equation}
F_{1}(R \gg 1/n) \approx \left (\frac{l_{c}}{R}\right )^{\frac{cm}{2\pi n \hbar}}
      \label{41p} \end{equation}
was found. Here, $c$ is the sound velocity, and $l_{c}$ is the healing length
($l_{c}=\hbar/\sqrt{mn\nu_{0}}$ \cite{petrov2003,petrov2004}).
Since $c=\sqrt{n\nu_{0}/m}$ (under a weak coupling),
formula (\ref{41p}) coincides with (\ref{40}), where  $f_{2}(R)= 1/\pi\approx 0.32$.
It was shown \cite{pethick2008} that relation (\ref{41p}) yields the formula
$N_{k\neq 0}\sim 1/|k|^{1-\sqrt{\beta}/2}$, which agrees with (\ref{41}), (\ref{43}).
If we change  $2\pi l/L \rightarrow k$ in formula (\ref{38}) for $N_{k\neq 0}$,
we obtain a formula \cite{gn1964,rc1967} for $N_{k\neq 0}$ in a \textit{three-dimensional} system.
We also mention works \cite{popov1980,berkovich1989,mora2003}, where
formula (\ref{40}) with $f_{2}(R)\approx  0.33$ was gotten.
Let us write formula (\ref{40}) as $F_{1}(R) \approx C(\gamma)(nR)^{-\sqrt{\gamma}/2\pi}$ \cite{olshanii2009}
($\gamma=\beta \pi^{2}$). Then,  in the thermodynamic limit ($L=\infty$) for $\gamma =0.001$ we obtain $C\approx 1.018$,
which is in agreement with the values $C\approx 1.016; 1.02$ \cite{olshanii2009}, obtained by two other methods.
The additional summand $(2/3)(R/L)^{2}$ in $f_{2}$ (\ref{39}) was not obtained earlier.
As far as we see, the reason lies in the transition to the thermodynamic limit or in
a not quite accurate calculation of sums.
We determined numerically a solution for $F_{1}(R)$, by using formulae (2.3.1) and (2.3.2) in  \cite{petrov2003}
and formula (15.44) in \cite{pethick2008}.
As a result, for $N=10^{4}$--$10^{6}$ and $\beta =10^{-4}$--$10^{-2},$
we obtain formula (\ref{40}) with
\begin{equation}
f_{2}(R)\approx  \frac{0.5+2(R/L)^{2}}{3}
      \label{50n} \end{equation}
instead of $f_{2}(R)= 1/\pi$ \cite{petrov2003,pethick2008}.
Both formulae give the same value for $R=L/2$. But, for other $R,$ formula (\ref{50n})
describes the solution better (which is well evident for $\ln F_{1}(R)$).
The distinction between $f_{2}(R)$ (\ref{50n}) and (\ref{39})
is apparently related to the fact that formulae \cite{petrov2003,pethick2008} were obtained in the low-energy approximation
($\epsilon (k)<\mu$), whereas our method involves all $k$.

Thus, the term $(2/3)(R/L)^{2}$  in $f_{2}(R)$ is a new result.
Its principal meaning consists in that, for the ground state, the decay law of the
density matrix turns out to be a not quite power one.
In addition, new results are formula (\ref{38}) for $N_{0}$ and the constant
in formula (\ref{38}) for $N_{k\neq 0}$.

 It was noticed \cite{petrov2004,pethick2008} that, for impenetrable point bosons, the substitution of the value
$cm/(2\pi n \hbar)=1/2$ \cite{girardeau1960}  into (\ref{41p}) gives
the proper formula $F_{1}(R)\sim R^{-1/2}$. On this basis, formula
(\ref{18c}) (without the constant) was deduced in
\cite{pethick2008}. However, the derivation of formula (\ref{41p})
in works
\cite{schwartz1977,haldane1981,petrov2003,petrov2004,pethick2008} is
valid only for a weak coupling. Indeed, the corrections to the
logarithm of the density matrix were not taken into account in
\cite{schwartz1977}, but these corrections are large for the strong
coupling. It follows from the formulae \cite{petrov2003,petrov2004}
that the fluctuations of a phase are connected with the fluctuations
of a concentration ($\delta \hat{n}$), and the smallness of $\delta
\hat{n}$ requires that $\nabla \hat{\varphi }(x)$ be small.
Moreover, the smallness of $\langle (\hat{\varphi }(x)-\hat{\varphi
}(0))^{2}\rangle = \sqrt{\beta}\ln{(x/l_{c})}$
\cite{petrov2003,pethick2008} for $x \sim l_{c}$ means the smallness
of $\beta$. By taking into account  only the long-wave fluctuations
of a phase, it was obtained for the strong coupling
\cite{petrov2004} $\langle (\hat{\varphi }(x)-\hat{\varphi
}(0))^{2}\rangle \simeq \ln{(x/l_{c})};$ here, the fluctuations are
not small. Therefore, such method is approximate.

Thus, formula (\ref{41p}) is valid also for a strong coupling.  The
possible reason consists in that only the two-particle correlations
are of importance for the ground state of the system for both strong
and weak couplings. Under an intermediate coupling, the higher
correlations are significant as well (sums with $a_{3}$, $a_{4},$
etc. in $\Psi_{0}$ (\ref{31})), and their consideration can change
the function (\ref{41p}).

In the two- and three-dimensional cases, we can analogously obtain for low levels
under a weak interaction:
\begin{eqnarray}
N^{2D}_{\textbf{k}\neq 0} \approx  N_{0}\frac{\sqrt{\beta_{2D}}}{4N^{1/2}\sqrt{j_{x}^{2}+j_{y}^{2}}},
\quad \beta_{2D}=\frac{\nu_{0}m}{\pi^{2}\hbar^{2}},
      \label{52} \end{eqnarray}
\begin{eqnarray}
N^{3D}_{\textbf{k}\neq 0} \approx  N_{0}\frac{\sqrt{\beta_{3D}}}{4N^{2/3}\sqrt{j_{x}^{2}+j_{y}^{2}+j_{z}^{2}}},
\quad \beta_{3D}=\frac{\nu_{0}mn^{1/3}}{\pi^{2}\hbar^{2}},
      \label{53} \end{eqnarray}
where
$\textbf{k}=2\pi(\frac{j_{x}}{L},\frac{j_{xy}}{L},\frac{j_{z}}{L})$,
$\nu_{0}=\int U(\textbf{r})d\textbf{r}$, $n=N/V$,
$L_{x}=L_{y}=L_{z}$. Thus, for large $N$ there are no
macroscopically filled levels with $\textbf{k}\neq 0$. The plausible
reason for this consists in a much larger number of a one-particle
levels as compared with the 1D case.

 \section{Possible experiment}
It is of interest whether it is possible for quasi-1D gases in a trap to enter into
the region $\beta \sim 0.01\div 0.1$.
In this case, we would be able to reveal experimentally quasicondensates on low levels.
We will make some estimates, by using the following parameters of a trap
\cite{druten2008}: $^{87}$Rb atoms ($a_{s}\approx 48\,\mbox{\AA}$ \cite{pethick2008}),
$N=2\cdot 10^7$, $\omega_{\rho}=2\pi\cdot 3280\,Hz$,  $\omega_{z}=2\pi\cdot 8.5\,Hz$,
$R_{z}\approx 0.54\,\mbox{mm}$, $R_{\rho}\approx 1.4\cdot 10^{-3}\,\mbox{mm}$, and
$T \gsim 10^{-7}\,K$. Since
$g_{1D}\equiv\nu_{1D}(0)=\frac{2\hbar^{2}a_{s}}{\mu a_{\rho}(a_{\rho}+\zeta(1/2) a_{s})}$
\cite{olshanii1998} ($\mu=m/2$ is the reduced mass,
$a_{\rho}=\sqrt{\hbar/\mu\omega_{\rho}}\approx 2600\,\mbox{\AA}$, $\zeta(1/2)\approx -1.46$), we obtain
\begin{equation}
\beta=\frac{\nu_{1D}(0)m}{\pi^{2}\hbar^{2}n} \approx
\frac{4a_{s}}{\pi^{2}na_{\rho}(a_{\rho}+\zeta(1/2)a_{s})}\approx 1.6\cdot 10^{-6}.
      \label{45} \end{equation}
For the experiment in \cite{aspect2003}, we obtain  from (\ref{45}) $\beta \simeq 4\cdot 10^{-5}$.
For the crude estimate of $N_{k},$ we use formulae (\ref{38})
deduced for a uniform system at $T=0$
(for $T > 0,$ the density matrix is multiplied by the factor  $\exp{[q_{1}(T)-q_{2}(T)R]}$ \cite{popov1980,schwartz1977,mora2003}
(with $q_{1}(T\rightarrow 0)\rightarrow 0$, $q_{2}(T)=\frac{mk_{B}T}{2\hbar^{2}n}$), which is close to 1
for $T\ll T_{f}=\frac{2\hbar^{2}n}{mk_{B}R_{z}}\simeq 4\cdot 10^{-7}\,K$
and has no influence on $N_{k}$). Then,
for $\beta$ (\ref{45}) and $N=2\cdot 10^7,$  relation  (\ref{38}) yields $N_{0}\approx 0.994N$. That is,
practically all atoms are in the condensate on the low level.
In this case, relation (\ref{39}) yields $|u_{1}(R)|\lsim 0.007$, $F_{1}(R)\approx const$,
i.e., the condensate is close to the true one.

 For a nonuniform gas in a trap, the eigenfunctions $f_{j}(x)$
are not plane waves, but are determined from the equation \cite{leggett2006}
\begin{equation}
F_{1}(x_{a},x_{b}) = \sum\limits_{j=0}^{\infty}\lambda_{j}f_{j}(x_{a})f_{j}^{*}(x_{b}).
\label{46}  \end{equation}
We can establish the connection between $N_{k}$ and $\lambda_{j}$:
\begin{equation}
\frac{N_{k}}{N} =\frac{1}{L^2}\int\limits_{-L/2}^{L/2}dx_{a}
\int\limits_{-L/2}^{L/2}dx_{b}F_{1}(x_{a},x_{b})e^{-ik(x_{a}-x_{b})}=\sum\limits_{j=0}^{\infty}\lambda_{j}|\chi_{j}(k)|^{2},
      \label{47} \end{equation}
\begin{equation}
\chi_{j}(k) =\frac{1}{L}\int\limits_{-L/2}^{L/2}dx f_{j}(x)e^{-ikx},
      \label{48} \end{equation}
where $k\equiv k_{l}=2\pi l/L$. It is of interest that, for impenetrable point bosons in a trap
at $T=0,$ the values of $\lambda_{j}$ \cite{forrester2003}
for $j=0, 1, 2,$  are close to the values of $N_{k_{j}}/N$ for the same uniform system.
This is related to the absence of a quasicondensate and to the fact that, for the given $l,$
$|\chi_{l}(k_{l})|$ is maximal among $|\chi_{j}(k_{l})|$.
In other words, for the strong coupling, \textit{the values of $\lambda_{j}$ for low levels of a uniform
system and a system in a trap are close, and the same is possible for a weak coupling.}
The system of bosons with a weak coupling in  a trap should contain a quasicondensate.
If almost all atoms are on the level $j=0$, then relation (\ref{47}) is reduced to
$N_{k_{l}}/N\approx \lambda_{0}|\chi_{0}(k_{l})|^{2}$. If a quasicondensate is present on several
levels, then all these levels $j$ should be taken into account in (\ref{47}).
For the systems considered in \cite{druten2008,aspect2003}, we have $\beta \ll 0.01$. Therefore, the condensate on the level $j=0$
contains, probably, almost all atoms. This means that
$F_{1}(x_{a},x_{b}) \approx \lambda_{0}f_{0}(x_{a})f_{0}^{*}(x_{b})$.
In this case, by the measured values of $N_{k}$
\cite{druten2012}, it is possible to restore $f_{0}(x)$ by the relations
$N_{k} =N\lambda_{0}|\chi_{0}(k)|^{2}$ and $f_{j}(x)=\sum^{(2\pi)}_{k}\chi_{j}(k)e^{ikx}$. Moreover,
the normalization conditions (\ref{29}), $\int^{L/2}_{-L/2}dxF_{1}(x,x)=L,$ and
$\int^{L/2}_{-L/2}dxf^{*}_{l}(x)f_{j}(x)=L\delta_{l,j}$  yield $\lambda_{0}\approx 1$.

It is seen from formula (\ref{45}) that the region $\beta \sim 0.01\div 0.1$ can be realized experimentally,
by varying $a_{s}$ by means of the Feshbach resonance \cite{feshbach,pit2006}.
For such $\beta,$ the number of atoms for the states with the  smallest $j$
should be macroscopic for each $j$ (analogous result was derived in \cite{witkowska2011};
we failed to determine $\beta$ by data \cite{witkowska2011}). In this case, the distribution
$N_{k}$ must be essentially different from $N_{k}$ for $\beta \ll 0.01$
(where almost all particles occupy the low level, and $\lambda_{0}\approx 1$).
The fragmented quasicondensate can be discovered in the following way: one needs
to measure $N_{k}$ \cite{druten2012,aspect2003,bouchoule2012} for $\beta \ll 0.01$ and
to restore $f_{0}(x)$  by $N_{k}$ in the above-described way. If this function turns out close to
$f_{0}(x)$ for impenetrable bosons in a trap \cite{forrester2003}, then we can use
the whole set $f_{j}(x)$ for such bosons and to find the whole set
$\lambda_{j}$ for bosons with $\beta \sim 0.1\div 0.01$ from (\ref{47}) and
the experimental value of $N_{k}$ for $\beta \sim 0.1\div 0.01$. If $f_{0}(x)$
for $\beta \ll 0.01$ would turn out to be considerably different from
$f_{0}(x)$ for impenetrable bosons \cite{forrester2003}, then we need to determine the density matrix for a system with
$\beta \sim 0.1\div 0.01$ and, by it, to calculate $f_{j}(x)$.
Then, from the experimental values of $N_{k},$ we can find $\lambda_{j}$.
The observation of a fragmented quasicondensate would be of interest.

 \section{Conclusion}
Using the formulae for the density matrix  \cite{vak1989,vak1990},
we have determined the average number of atoms  with momentum $\hbar k$
on low levels for the ground state
of a one-dimensional uniform periodic system of interacting bosons.
The solutions agree with previously obtained ones.
The new results are as follows: For impenetrable point bosons, the solution in the second approximation is found
(earlier, only a solution in the first approximation was obtained).
For almost point bosons with weak coupling, we deduced the formula for $N_{0}$ and made
the formula  for the density matrix $F_{1}(R)$ to be somewhat more accurate.
The most interesting result consists in the finding that the uniform
system of bosons with weak coupling
can possess the  quasicondensate on many low levels. Such fragmented
1D-quasicondensate can be investigated by the use of a gas in a trap.

     \renewcommand\refname{}


       \end{document}